\documentclass[pra,twocolumn,superscriptaddress,notitlepage,longbibliography]{revtex4-2}
\usepackage{amsmath}
\usepackage{amssymb}
\usepackage{bbold}
\usepackage{natbib}
\usepackage[unicode=true,colorlinks=true,citecolor=blue,urlcolor=blue]{hyperref}

\usepackage{bm}
\usepackage{epsfig}
\usepackage{tikz}
\usepackage[normalem]{ulem}

\renewcommand {\Im}{\mathop\mathrm{Im}\nolimits}
\renewcommand {\Re}{\mathop\mathrm{Re}\nolimits}
\newcommand {\Tr}{\mathop\mathrm{Tr}\nolimits}
\renewcommand {\phi}{{\varphi}}
\newcommand {\rmi}{{\rm i}}

\newcommand {\e}{{\rm e}}

\begin{document}
\title{Resonant Parametric Photon Generation\\ in Waveguide-coupled Quantum Emitter Arrays
}

\author{Egor S. Vyatkin}
\affiliation{Ioffe Institute, St. Petersburg 194021, Russia}

\author{Alexander V. Poshakinskiy}
\affiliation{Ioffe Institute, St. Petersburg 194021, Russia}

\author{Alexander N. Poddubny}
\affiliation{Department of Physics of Complex Systems, Weizmann Institute of Science, Rehovot 7610001, Israel}

\email{poddubny@weizmann.ac.il}

\begin{abstract}
We have developed a theory of  parametric photon generation in the waveguides coupled to arrays of quantum emitters with temporally modulated resonance frequencies. Such generation can be interpreted as a dynamical Casimir effect. 
We demonstrate numerically and analytically how the emission directionality and photon-photon correlations  can be controlled by the phases of the modulation.  The emission spectrum is shown to be strongly dependent on the  anharmonicity of the emitter potential. Single- and double-excited state resonances have been identified in the  emission spectrum.
\end{abstract}
\date{\today}

\maketitle

\section{Introduction}\label{sec:intro}

Waveguide quantum electrodynamics, describing photon interaction with arrays of emitters coupled to the waveguide, is now rapidly developing~\cite{sheremet2021waveguide,Roy2017,KimbleRMP2018}. This platform allows controllable generation of quantum light \cite{Prasad2020}, and control over 
lifetimes~\cite{zanner2021coherent} and entanglement
\cite{Kannan2020,Corzo2019} of coupled atom-photon excitations.  
Even more possibilities are opened in the structures with the parameters dynamically modulated in time~\cite{Silveri_2017}. This enables  Floquet engineering and realization of synthetic dimensions~\cite{Ozawa2019,Yuan:18} as well as   control of quantum photon-photon correlations~\cite{Ilin2022,Haviland2021}. Such time modulation has been recently demonstrated, for example, for  the superconducting transmon emitter platform ~\cite{Redchenko2022}. 

One more fundamental physical effect, that becomes possible in the 
dynamically modulated structures, is the parametric generation of photon pairs. Such generation can be also interpreted as a dynamical Casimir effect. This effect has been first proposed for the cavity with a moving wall~\cite{Moore1970}. To the best of our knowledge, it has been  never directly observed in this setup so far, because 
of the extremely low photon generation rate at realistic parameters, see the review ~\cite{Dodonov}. However, there also exist generalized dynamical Casimir effects, where other electromagnetic properties of the medium are changing instead of physical movement of the mirror in space. For example, parametric photon generation due to electron-hole plasma generated and moving in a semiconductor under the laser pulse excitation has been theoretically considered in Ref.~\cite{Lozovik1995}. A seminal observation of an analogue of the dynamical Casimir effect in a superconducting circuit was made in Ref.~\cite{Wilson2011}. The effective length of the transmission line has been modulated  by changing the inductance of a superconducting quantum interference device. A detailed theoretical analysis of such systems has been performed in Refs.~\cite{Nori2010,Nori2018}.  
Two-photon entanglement in this setup has been experimentally studied in Ref.~\cite{Schneider2020}.
However, the consideration has been limited to the case when the modulated elements were not resonant for the generated photons. Dynamical Casimir effect in the arrays of resonant modulated emitters is not yet explored. The resonant structures could potentially allow one to selectively enhance the photon generation and control the correlations between them. Recently, a theory of dynamical Casimir effect in the general dynamically modulated photonic structures has been put forward in Ref.~\cite{Sloan2021}. However, considered setup  did not include two-photon interactions, essential for the emitter platform.

Here, we consider  a parametric photon generation by an array of emitters with strongly dynamically modulated resonant frequencies, coupled to the waveguide. In such a system there exist resonances for generated photons and photon pairs near the single- and double-excited levels of the emitters. Our goal is to explore the role of the various collective emitter resonances for the intensity, directionality and the quantum correlations between the emitted photons.

\begin{figure}[b!]
    \centering
    \includegraphics[width=0.95\columnwidth]{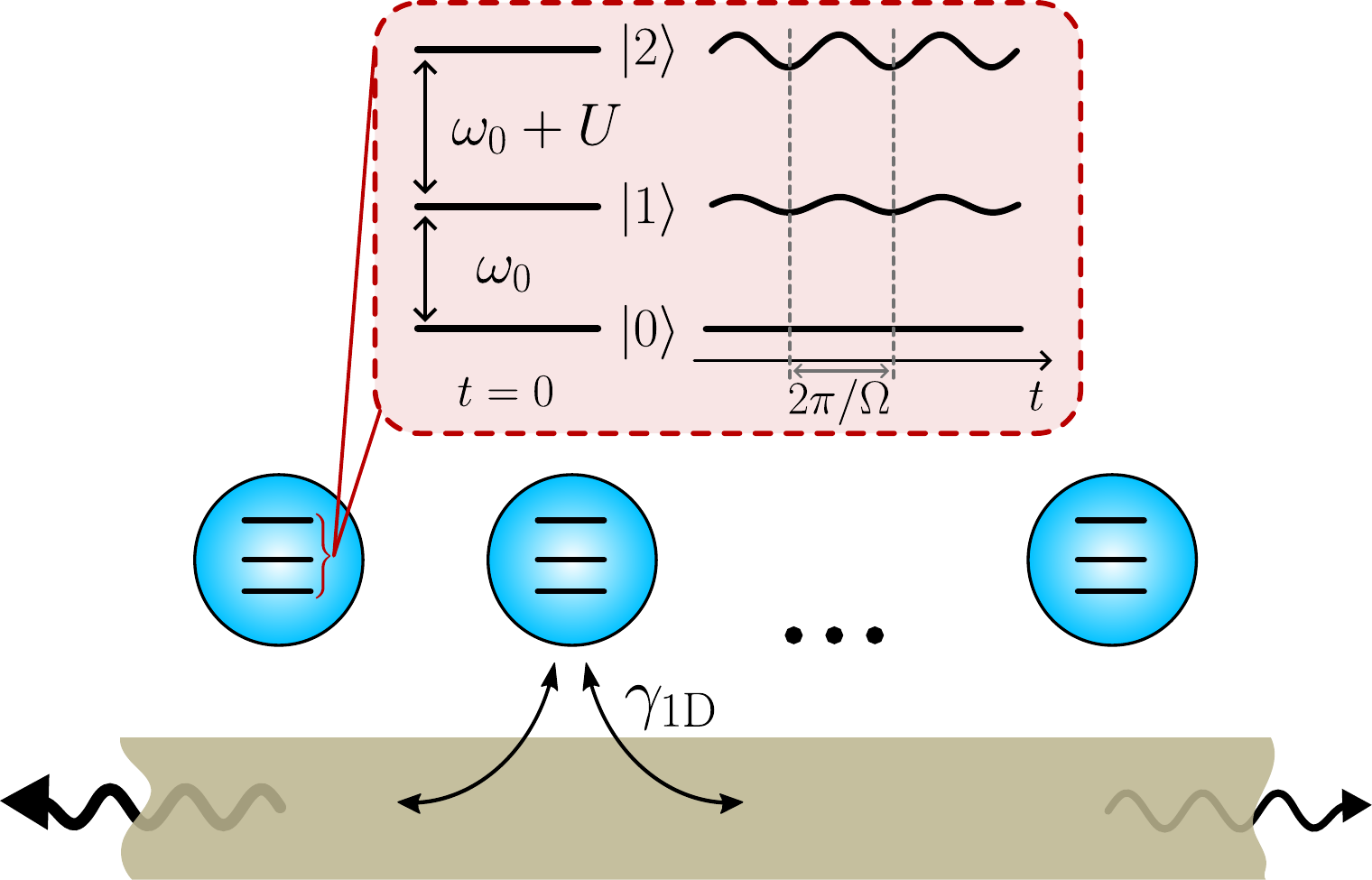}
    \caption{Scheme of the structure under consideration. The array of emitters with modulated in time resonant frequencies is coupled to a waveguide.}
    \label{fig:scheme}
\end{figure}

The rest of the paper is organized as follows. Our theoretical model and calculation approach, based on the master equation, are presented in Sec.~\ref{sec:model}. We discuss the numerical and analytical results for a single emitter in Sec.~\ref{sec:single}. The arrays with $N>1$ emitters are considered in Sec.~\ref{sec:arrays} and the main results are summarized in Sec.~\ref{sec:summary}. Appendix~\ref{app:diagrams} presents an equivalent alternative Green-function-based diagrammatic approach to calculate the photon emission intensity and the photon-photon correlation functions. Analytic results obtained by this approach for a particular case of $N=2$ emitters are given in Appendix~\ref{app:two}.




\section{Model}\label{sec:model}
The structure under consideration is schematically illustrated in Fig.~\ref{fig:scheme}. It consists of $N$ periodically spaced emitters in the one-dimensional waveguide. The resonant frequencies of the emitters are modulated in time with the modulation frequency $\Omega$. The Hermitian part of the system Hamiltonian can be written as $H=H_0+\sum_j^N V_j$, where ($\hbar=1$)
\begin{equation}\label{eq:H0}
    H_0=\sum_{j=1}^N\left(\omega_0 a_j^\dagger a_j+\frac{U}{2}a_j^\dagger a_j^\dagger a_j a_j\right)+\sum_{j,k=1}^N\Re(D_{jk})a_j^\dagger a_k
\end{equation}
is the unperturbed Hamiltonian and $a_j$ are
the bosonic annihilation operators for the emitter excitations, located at the point $x_j=d(j-1)$.
The first term of Eq.~\eqref{eq:H0} describes the structure of the emitter energy levels which is shown in the red frame in Fig.~\ref{fig:scheme} (first three levels) for $t=0$. The energy of the first excited level is $\omega_0$ while the second excited level has energy $2\omega_0+U$ assuming the anharmonicity $U\ll\omega_0$. The radiative coupling and collective decay of the emitters are described by the photon Green function
\begin{equation}
    D_{jk}=-\rmi \gamma_{1\rm{D}} e^{\rmi \omega_0|x_j-x_k|/c}\;,
\end{equation}
where $\gamma_{1\rm{D}}$ is the  radiative decay rate of a single emitter. The  coupling is long-ranged since it is mediated by photons, propagating into the waveguide~\cite{sheremet2021waveguide}. The perturbation responsible for modulating the frequency of each emitter reads  
\begin{equation}\label{eq:V}
    V_j=g_j(a_j^\dagger+a_j)^2\cos(\Omega t+\varphi_j)\;.
\end{equation}
Here, $g_j$ and $\phi_j$ are the amplitude and the phase of the modulation, respectively. 
The time evolution of the system is described by the master equation~\cite{sheremet2021waveguide,Ilin2022}
\begin{align}\label{eq:master}
    \dot{\rho}=-\rmi[H,\rho]+\sum_{j,k=1}^N\gamma_{jk}(2 a_j \rho a_k^\dagger-a_j ^\dagger a_k\rho-\rho a_j^\dagger a_k)\;,
\end{align}
where $\gamma_{jk}=-\Im(D_{jk})+\delta_{jk}\gamma$ and $\gamma$ is the nonradiative decay rate. Such system, with the anharmonicity term and the parametric driving, can be experimentally realized in the microwave spectral range by superconducting quantum $LC$ circuits, where the  Josephson junctions are used as nonlinear inductance~\cite{Naeini2019}. Such systems are now widely used as superconducting qubits, see the review~\cite{Blais2020} and their parameters satisfy the regime
$\omega_0\gg U\gg \gamma_{\rm 1D}\gg \gamma$ considered here. For example, the parameters of Ref.~\cite{zanner2021coherent} correspond to $\omega_0\approx 7~$GHz, $U\sim 0.2~$GHz, $\gamma_{\rm 1D}\approx 0.03~$GHz, $\gamma\sim 10^{-3}$GHz.

\begin{figure}[b!]
\centering{\includegraphics[width=0.9\columnwidth]{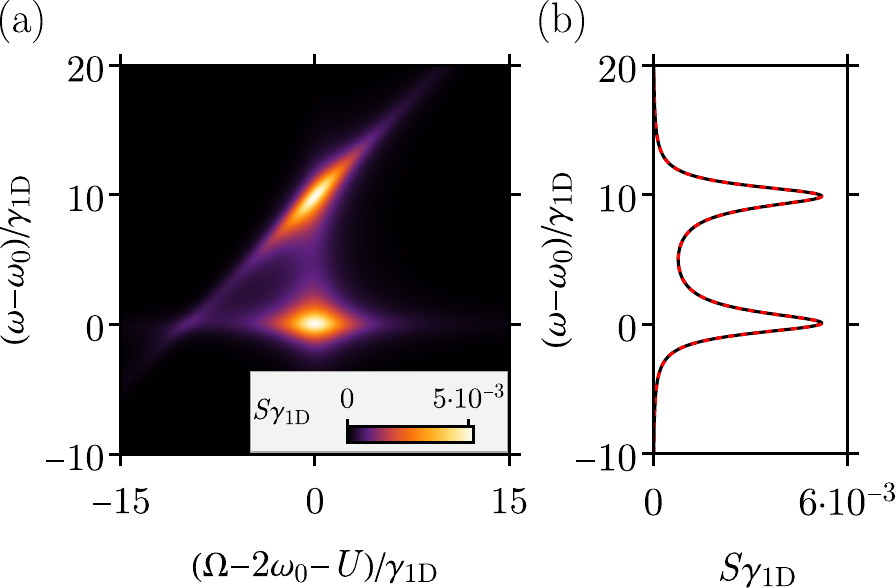}}
\caption{(a) Emission spectrum $S$ of one emitter as a function of the modulation frequency $\Omega$ and the emission frequency $\omega$. Spectrum is calculated using the density matrix method described in Appendix~\ref{app:Spectrum}. The calculation parameters are $\omega_0/\gamma_{1\rm{D}}=200,\;g/\gamma_{1\rm{D}}=0.1,\;U/\gamma_{1\rm{D}}=10,\;\gamma=0$. The black line on the panel (b) shows the emission spectrum
obtained as the cross section of the density plot (a) for $\Omega=2\omega_0+U$ and the red dashed line shows the spectrum for the same parameters obtained using the diagrammatic approach.}
\label{fig:1-spectr}
\end{figure}
\section{Single emitter}\label{sec:single}

We start by considering a single emitter coupled to a waveguide. We are interested in the weak driving regime when $g\ll\gamma_{1\rm{D}}$. In this case, to find the observable quantities, it is sufficient to restrict the consideration to a three-level emitter.
The method of finding the stationary density matrix (solution of Eq.~\eqref{eq:master} for $\gamma_{\rm 1D}t\gg 1$) is described in Appendix~\ref{app:Master}. 
The total photon emission rate reads $2 \gamma_{1\rm{D}} I_1$, i.e., is determined by the number or emitter excitations   
\begin{align}\label{eq:I1}
    I_1& =  {\rm Tr}(\rho_0 a^\dagger a)\\ \nonumber
    &=\frac{4 g^2  (\Omega^2+(U+2\omega_0)^2+4\gamma_\Sigma^2)}{[4\gamma_\Sigma^2 + (\Omega-U-2\omega_0)^2] [4\gamma_\Sigma^2+(\Omega + U+2\omega_0)^2]}\;,
\end{align}
and their radiative decay rate. Here, $\rho_0$ is time-averaged density matrix of the emitter and $\gamma_\Sigma=\gamma_{1\rm{D}}+\gamma$ (see Appendix~\ref{app:Master-1qub}). The highest intensity is achieved when the modulation frequency is in resonance with the transition between the ground level and the second excited level of the emitter, $\Omega = 2\omega_0 +U$. 

Another important characteristic is the spectrum of the photon emission. It can be found using the
quantum regression theorem as described in Appendix~\ref{app:Spectrum}. The result of the calculation according to Eq.~\eqref{eq:spectr-master} is shown in Fig.~\ref{fig:1-spectr}
 As mentioned above, the maximum integral intensity is reached at the modulation frequency $\Omega = 2\omega_0 +U$. The emission spectrum consists of two peaks, one at the  frequency $\omega_0$, which corresponds to the transition between the ground and first excited levels of the emitter, and the other peak at the frequency $\Omega-\omega_0$. This reflects the fact that the photons are born and emitted in pairs with the average energies $\omega_0$ and $\Omega-\omega_0$,  i.e. with the total energy $\Omega$. 
The spectrum for the resonant modulation frequency obtained with the density method is shown by the black line in panel (b). The diagrammatic approach (Appendix~\ref{app:diagrams}) yields the spectrum
 \begin{align}\label{eq:spectr}
    S(\omega)=&\frac{2g^2\gamma_{\rm 1D}}{\left[(\Omega-2\omega_0-U)^2+4\gamma_{\rm 1D}^2\right]}\\ \nonumber
    &\times\frac{\left[(\Omega-2\omega_0)^2+4\gamma_{1\rm D}^2\right]}{\left[(\omega-\omega_0)^2+\gamma_{\rm 1D}^2\right]\left[(\omega-\Omega+\omega_0)^2+\gamma_{\rm 1 D}^2\right]}\;,
\end{align}
which is shown by the red dashed line and as we can see it  perfectly agrees with the black line.
\begin{figure}[t!]
    \centering
    \includegraphics[width=\columnwidth]{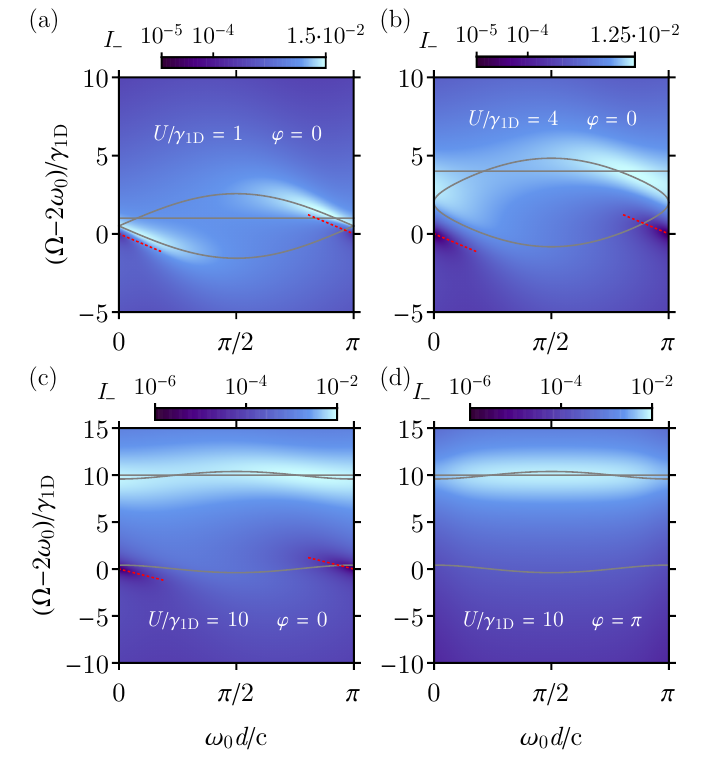}
    \caption{The photon emission intensity to the left $I_-=\langle p_-^\dagger p_-\rangle$ for two emitters as a function of a distance between the emitters $d$ and the modulation frequency $\Omega$. Calculations have been performed in the case of symmetric modulation ($\varphi=0$) for (a) $U/\gamma_{1\rm{D}}=1$, (b) $U/\gamma_{1\rm{D}}=4$, (c) $U/\gamma_{1\rm{D}}=10$ and (d) antisymmetric modulation ($\varphi=\pi$) for $U/\gamma_{1\rm{D}}=10$. Other calculation parameters are $\omega_0/\gamma_{1\rm{D}}=200, g/\gamma_{1\rm{D}}=0.1, \gamma/\gamma_{1\rm{D}}=0.1$. Gray lines represent the double-excited eigenstates. Red dotted lines show the intensity minima.}
\label{fig:Intens}
\end{figure}

\begin{figure}[t!]
    \centering
    \includegraphics[width=\columnwidth]{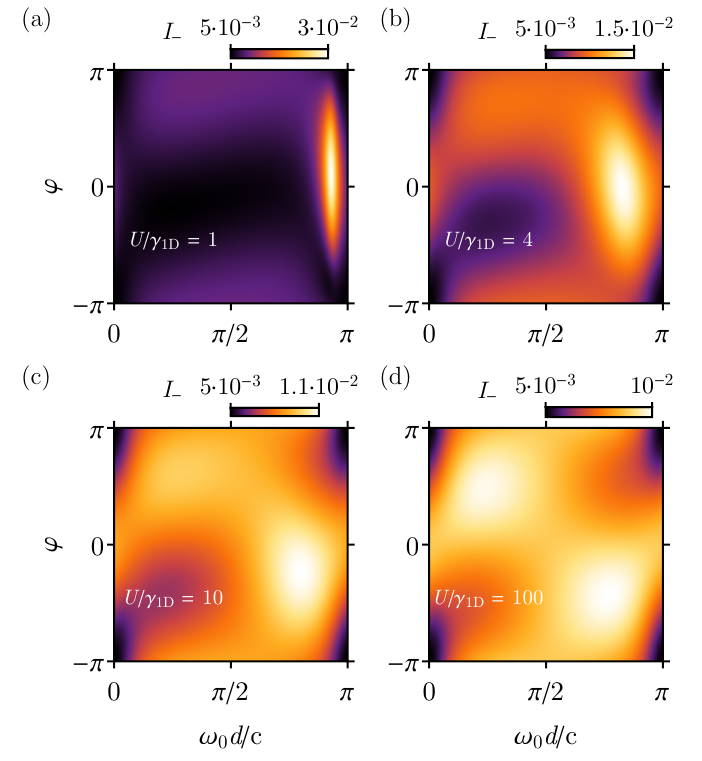}
    \caption{The photon emission intensity to the left $I_-$ for a pair of emitters as a function of the distance $d$ and the modulation phase $\varphi$ for the modulation frequency $\Omega=2 \omega_0 + U$ and different $U$ indicated in the figure. Calculation has been performed for $g/\gamma_{1\rm D}=0.1$, $\omega_0/\gamma_{1\rm{D}}=200$ and $\gamma/\gamma_{1\rm D}=0.05$.  }
\label{fig:Orient}
\end{figure}

\section{Emitter arrays}\label{sec:arrays}
We now proceed to the discussion of the paramatric generation from emitter arrays.
\subsection{$N=2$ emitters}
We start with the case of just a pair of emitters. We  assume that the coupling constants are equal, $g_1=g_2=g$, and  that the  phases are $\varphi_1=0$, $\varphi_2=\varphi$. Unlike the case of a single emitter, this system can emit directionally  because of the interference between photons from the first and the second emitter. 

The photons travelling in the left (right) direction are coupled to the combination of the emitters lowering operators $p_{\mp}=a_1+a_2\e^{\pm \rmi q d}$, where $q d \equiv \omega_0 d/c$ is the phase gained by the photon as it travels between the two emitters. The photon emission intensity to the left is determined by $I_-={\rm Tr}(\rho_0  p_-^\dagger p_-)$~\cite{sheremet2021waveguide} and can be obtained by using the stationary density matrix (Appendix~\ref{app:Master}).

\begin{figure}[t]
    \centering
    \includegraphics[width=\columnwidth]{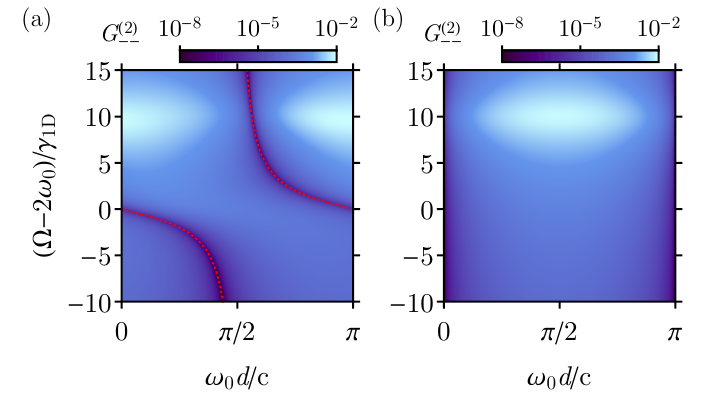}
    \caption{The second order correlation function for a pair of emitters in the case of (a) symmetric $(\varphi=0)$ and (b)  antisymmetric $(\varphi=\pi)$ modulation. Calculation has
been performed for the following set of parameters: $U/\gamma_{1\rm{D}}=10$, $\omega_0/\gamma_{1\rm{D}}=200$, $g/\gamma_{1\rm{D}}=0.1$, $\gamma/\gamma_{1\rm{D}}=0.1$.}
\label{fig:g2}
\end{figure}

Figure \ref{fig:Intens} presents the calculation of $I_-$ for the three values of the parameter $U/\gamma_{1\rm D}$. Panels (a), (b) and (c) correspond to the case of symmetric modulation, when both emitters are excited in phase ($\varphi=0$). In this case, we can see that the sharp intensity minima appear in the vicinity of red dotted lines which correspond to dark single-excited states. The highest intensity is achieved when the modulation frequency is close to the real parts of the energies of the certain double-excited eigenstates (gray lines). The latter are found by diagonalizing the non-Hermitian two-photon Hamiltonian $H\otimes 1+1\otimes H+\mathcal{U}$, where $H$ and $\mathcal{U}$ are defined in Appendix~\ref{app:diagrams}. For in-phase modulation the intensity is enhanced near symmetric states in terms of emitter permutation, whose energies are
\begin{align}
    \varepsilon_{1,2}=2\omega_0-2\rmi \gamma_{1\rm D}+\frac{U}{2}\pm\frac{1}{2}\sqrt{U^2-16\gamma_{1\rm D}^2 e^{2\rmi q d}}
\end{align}
If $U/\gamma_{1\rm D}<4$ [panel (a)], the energies of the three eigenstates are close and intersect when $\omega_0d/c$ is varied. As $U/\gamma_{1\rm D}$ is increased, at a certain threshold [panel (b), $U/\gamma_{1\rm D}=4$] the symmetric states are rearranged into two isolated bands, and that get separated by the gap of $U$ for large $U/\gamma_{1\rm D}$ [panel (c)]. 
The upper state $\approx (a_1^\dagger a_1^\dagger+a_2^\dagger a_2^\dagger)|0\rangle$ has the energy $\Re\varepsilon_1\approx 2\omega_0+U$ which is almost independent of the distance $d$, but the lower band (with the energy $\Re\varepsilon_2\approx 2\omega_0$) is mainly formed by the states that comprise pair of excitations in different emitters, $a_1^\dag a_2^\dag|0\rangle$. 
Such states cannot be excited by the perturbation operator Eq.~\eqref{eq:V}, that includes the products of operators two $a_j^\dag$ with the same $j$ only.  
Therefore, the emission intensity near the lower band is quenched, see also Appendix \ref{app:two} for the complete analytical expression. The case of out of phase modulation ($\varphi =\pi$) is shown in the panel (d).  
Here, we see no minima at the frequency of dark single-excited states and the overall dependence on distance $d$ is rather weak, because the energy of antisymmetric state $(a_1^\dagger a_1^\dagger-a_2^\dagger a_2^\dagger)|0\rangle$ does not depend on the distance and is equal to $2\omega_0+U$.  The maximum intensity is observed near this frequency.

The dependence of the emission on the modulation phase is shown in more detail in Fig.~\ref{fig:Orient}, which presents the calculation of $I_-$ for the modulation frequency $\Omega=2 \omega_0 + U$ that corresponds to the  resonance for a single emitter. In the case of $U/\gamma_{1\rm D}=1$ [panel (a)] the maximum is observed at $\omega_0 d/c \approx \pi$ which corresponds to the  resonance with a double-excited eigenstate, cf. Fig.~\ref{fig:Intens}(a). As the ratio $U/\gamma_{1\rm D}$ increases, the structure of $I_-$ becomes more smooth along the $d$ axis, due to the decreasing dependence of the eigenstates on the distance $d$. In contrast, the dependence of $I_-$ on $\varphi$ becomes more pronounced, see 
Fig.~\ref{fig:Intens}(b)-(d).
Note that the photon emission intensity to the right $I_+$ can be obtained by flipping the sign of $\varphi$. The obtained maps, especially Fig.~\ref{fig:Intens}(b), are highly asymmetric with respect to this operation, which indicates the high directivity of the emission.
In the case $U/\gamma_{\rm 1D}\gg 1$, which is shown in panel (d), the answer can be obtained analytically. At $\Omega= 2\omega_0+U$, we can neglect the state when different emitters are excited simultaneously. Then, we find 
\begin{equation}\label{eq:Iminus}
    I_-=I_1\left[2+\frac{\sin \varphi\sin(2\omega_0 d/c)}{3-\cos(2\omega_0 d/c)}\right]
\end{equation}
in the absence of the nonradiative decay rate $\gamma=0$, see also Appendix~\ref{app:two}. 
The second term in the right-hand side of Eq.~\eqref{eq:Iminus} describes the interference between the photon pairs generated by the two emitters.
Importantly, the interference term is odd in $\varphi$, thus its contribution is opposite for $I_-$ and $I_+$ which results in a directional emission. The maximal degree of directivity  $(I_--I_+)/(I_-+I_+) =\sqrt{2}/8 \approx 0.18$ is achieved for  $\varphi=\pi/2$ and $\omega_0 d/c = \arctan (2\sqrt{2})/2\approx\pi/5$. The intensity reduction in the corners of the panels in Fig.~\ref{fig:Orient} is out of the scope of Eq.~\eqref{eq:Iminus} and explained by  the finite nonradiative decay that was taken into account in the calculation.

We also present in Fig.~\ref{fig:g2} the (unnormalized)   photon-photon correlation function 
$G^{(2)}_{--}=\langle p_-^\dagger p_-^\dagger p_- p_-\rangle$
for symmetric [panel (a)] and antisymmetric [panel (b)] modulation of the emitters. The calculation demonstrates that the largest values of $G^{(2)}_{--}$ are
achieved at the two-photon resonance $\Omega=2\omega_0+U$.  In order  to explain the calculated dependence of the correlation function on the distance between the emitters  $d$ we have obtained an analytical expression similar to Eq.~\eqref{eq:Iminus} and valid  for $U/\gamma_{1\rm D}\gg1$\:,
\begin{equation}\label{eq:g2mm}
    G^{(2)}_{--}= I_1\left[1+\cos(2\omega_0 d/c-\varphi)\right]\:,
\end{equation}
see also Appendix~\ref{app:two}. The simple form of Eq.~\eqref{eq:g2mm} describes the interference of the two independent coherent sources separated by distance $d$ and emitting with the phase difference  $\varphi$. The fact that the two-photon emission and detection is considered is accounted by the factor of 2 in the phase $2\omega_0 d/c$ that the photon pair gains when travelling the distance $d$.
If the emitters are modulated in phase ($\varphi=0$), the maxima of Eq.~\eqref{eq:g2mm} are realized for the periods when $\omega_0 d/c=0,\pi,2\pi$ etc. The out-of-phase modulation, when $\varphi=\pi$ corresponds to the maxima at $\omega_0 d/c=\pi/2,3\pi/2\ldots$. This agrees with the numerical calculations in Fig.~\ref{fig:g2}.
For symmetric modulation there exists also an additional minimum in Fig.~\ref{fig:g2}(a), corresponding to a strong antibunching. According to our analytical expression Eq.~\eqref{eq:g2symm}, this minimum corresponds to the frequency $\Omega = 2\omega_0 -2 \gamma_{\rm 1D} \tan qd 
$. The corresponding expression is shown by a red dotted line in Fig.~\ref{fig:g2}(a) and well describes numerical results.

\begin{figure}[t]
    \centering
    \includegraphics[width=\columnwidth]{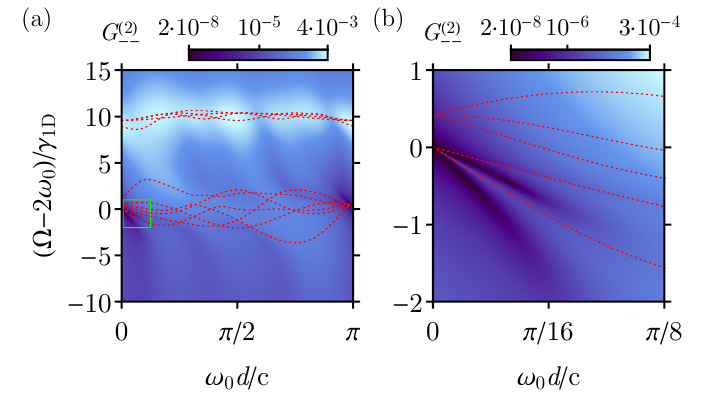}
    \caption{(a) The second order correlation function for an array of four emitters in the case of frequency modulation of the first emitter only. Panel (b) displays the area highlighted with a green square on the panel (a). Red dotted curves present the real parts of the energies of double-excited states. Calculation has been performed for the following set of parameters: $\omega_0/\gamma_{1\rm{D}}=200$, $U/\gamma_{1\rm{D}}=10$, $g/\gamma_{1\rm{D}}=0.1$, $\gamma/\gamma_{1\rm{D}}=0$.}
    \label{fig:4qubits}
\end{figure}

\subsection{$N=4$ emitters}
We have also calculated the two-photon correlation function for a larger array with $N=4$ emitters.  This is the minimum emitter number required  to have double-excited subradiant states~\cite{Ke2019,Molmer2019,zanner2021coherent}. Such states  have long radiative lifetime because of the destructive interference in the spontaneous photon emission processes. They exist for 
$\omega_0d/c\ll 1$ (or
$|\omega_0d/c-\pi|\ll 1$), and their lifetime is enhanced by the factors on the order of $1/(\omega_0d/c)^2$ ($1/|\omega_0d/c-\pi|^2$).
Hence, we can expect appearance of additional sharp spectral features in $G^{(2)}_{--}$ for $N=4$ due to the double-excited subradiant states. The corresponding color plots of calculated $G^{(2)}_{--}$ are shown in Fig.~\ref{fig:4qubits}. Correlation function $G^{(2)}_{--}$ has maxima around $\Omega=2\omega_0+U$, similarly to the case of $N=2$ emitters. As expected, there also appear  minima at $\Omega \approx 2\omega_0$ when the distance between the emitters is either small or close to $c\pi/\omega_0$. The map of $G^{(2)}_{--}$ in the region of small $\omega_0 d/c$ is shown in pannel (b) in detail. One can see that on top of the smooth minimum two sharp maxima appear [see two lower dotted lines in Fig.~\ref{fig:4qubits}(b)]. Their position matches the energies of the  two subradiant states $2\omega_0-2\gamma_{1\rm D}\omega_0 d/c$ and  $2\omega_0-(14/3)\gamma_{1\rm D}\omega_0 d/c$~\cite{Ke2019}.

\section{Summary}\label{sec:summary}
To summarize, we have developed a general theory of parametric photon generation from arrays of emitter coupled to the waveguide, that are modulated in time. 
Using the two independent approaches, the master equation for the density matrix and the diagrammatic Green-function technique, 
we have studied the dependence of the photon emission spectrum and photon-photon correlation functions on the anharmonicity of the emitter potential $U$, the distance between the neighboring emitters and the relative modulation phase $\varphi$.  

The calculated emission spectrum is very sensitive to the emitter anharmonicity parameter. The anharmonicity controls the relative weight of the spectral features around the single-  and  double-excited emitter resonances. The latter become more prominent with the increase of the anharmonicity.  When the number of emitters is $N=4$ or larger, additional sharp spectral features, corresponding to the double-excited subradiant states, appear in the spectrum.
We have also shown that the interference between photons emitted from different emitters can be controlled by relative phases of their frequency modulation. Our calculation demonstrates how this can be used to obtain directional photon pair emission, similarly as it happens for non-parametric quantum photon sources~\cite{Oliver2022,Redchenko2022,Golub2021}. 

 We hope that our results will  be useful for engineering the parametric quantum emission from the waveguide-coupled emitter arrays. A potentially interesting future research direction could be the system, where the spatial position of the light emitters, rather than their resonance frequency, oscillates in time. This would mean generalization of our concept of optomechanical Kerker effect~\cite{Kerker2019}, that is motion-induced direcitonal emission, to the quantum  optics regime.

\acknowledgements
We thank E.S. Redchenko for useful discussions.
This work has been funded by the Russian Science Foundation Grant No. 20-12-00194. 
Development of the diagrammatic approach was supported by the Russian Science Foundation Grant No. 21-72-10035.

\appendix

\section{Master equation approach}
\label{app:Master}

In order to find a stationary solution for the density matrix, we represent $\rho$ as a vector $\boldsymbol{\rho}$ according to the rule $\boldsymbol{\rho}_{n(i-1)+j}=\rho_{ij}$ for $n\times n$ matrix $\rho$. Then Eq.~\eqref{eq:master} can be written as
\begin{equation}\label{eq:vector_rho}
    \dot{\boldsymbol{\rho}}=\mathcal{L}\boldsymbol{\rho}+\frac{1}{2}\left(\mathcal{V} e^{\rmi\Omega t}-\mathcal{V}^*e^{-i\Omega t}\right)\boldsymbol{\rho}\;,
\end{equation}
where the operators $\mathcal{L}$ and $\mathcal{V}$ read
\begin{align}
    \mathcal{L}&=-\rmi(H_0\otimes 1-1\otimes H_0)\\ \nonumber
    &+\sum_{j,k=1}^N\gamma_{jk}(2 a_j\otimes a_k-a_j^\dagger a_k\otimes 1-1\otimes a_k^\dagger a_j)\;,\\ \nonumber
    \mathcal{V}&=-\rmi\sum_{j=1}^N g_j e^{\rmi \varphi_j}\left((a_j^\dagger+a_j)^2\otimes 1-1\otimes (a_j^\dagger+a_j)^2\right)\;,
\end{align}
and the symbol $\otimes$ denotes the Kronecker product. We solve Eq.~\eqref{eq:vector_rho} for $\gamma_{1\rm D}t\gg1$, assuming that at $t=0$ the modulation is turned on. Representing the density matrix in the form of a Fourier series $\boldsymbol{\rho}(t)=\sum_n e^{-\rmi n \Omega t}\boldsymbol{\rho}_n$, leaving only the harmonics with $n=-1,0,1$ and substituting it  into Eq.~\eqref{eq:vector_rho} we obtain  the harmonics with $n=\pm 1$ and a system of equations for the zero harmonic of the density matrix
\begin{align}\label{eq:rho_stat-1}
    \boldsymbol{\rho}_{-1}&=-\frac{1}{2}(\mathcal{L}-\rmi \Omega)^{-1}\mathcal{V}\boldsymbol{\rho}(0)\;,\\ \label{eq:rho_stat+1}
    \boldsymbol{\rho}_{1}&=\frac{1}{2}(\mathcal{L}+\rmi \Omega)^{-1}\mathcal{V}^*\boldsymbol{\rho}(0)\;,\\ \label{eq:rho_stat0}
    \mathcal{L}\boldsymbol{\rho}_0=-\frac{1}{4}&\left(\mathcal{V}\left(\mathcal{L}+\rmi \Omega \right)^{-1}\mathcal{V}^*+\mathcal{V}^*\left(\mathcal{L}-\rmi \Omega \right)^{-1}\mathcal{V}\right)\boldsymbol{\rho}(0)\;.
\end{align}
Here, a vector $\boldsymbol{\rho}(0)=\begin{pmatrix}
    1& 0 &\ldots & 0 
\end{pmatrix}$ corresponds to the initial vacuum state. 
\subsection{Emission of a single emitter} \label{app:Master-1qub}
 In weak modulation regime $g\ll\gamma_{\rm 1 D}$, we can restrict the consideration to only the first three levels of the emitter and take the annihilation operator in the form
 \begin{equation}
     a=\begin{pmatrix}
    0 & 1 & 0\\
    0 & 0 & \sqrt{2}\\
    0 & 0 & 0
    \end{pmatrix}\;.
 \end{equation}
 We get linear in $g$ harmonic with $n=-1$ from Eq.~\eqref{eq:rho_stat-1} with non-zero elements
\begin{align}
    (\rho_{-1})_{13}&=\frac{g}{\sqrt{2}\left[(\Omega-2\omega_0-U)-2\rmi \gamma_{\Sigma}\right]}\;,\\ \nonumber
     (\rho_{-1})_{31}&=-\frac{g}{\sqrt{2}\left[(\Omega+2\omega_0+U)-2\rmi \gamma_{\Sigma}\right]}\;,
\end{align}
and from Eq.~\eqref{eq:rho_stat+1} we get $\rho_1=\rho_{-1}^\dagger$. The solution of Eq.~\eqref{eq:rho_stat0} yields non-zero elements of the 
 quadratic in $g$ zero harmonic
\begin{align} \label{eq:rho0-1qub}
    (\rho_0)_{13}&=-\frac{g^2}{\sqrt{2}(\Omega^2+(2\gamma_\Sigma-\rmi (U+2\omega_0))^2)}\;,\\ \nonumber
    (\rho_0)_{22}&=2 g^2\\ \nonumber
    \times&\frac{(4\gamma^2_\Sigma+\Omega^2+(U+2\omega_0))}{\left[4\gamma_\Sigma^2+(\Omega-2\omega_0-U)^2\right]\left[4\gamma^2_\Sigma+(\Omega+2\omega_0+U)^2\right]}\;,\\ \nonumber
    &\;\;\;\;\;\;(\rho_0)_{31}=(\rho_0)_{13}^*\;,\;\;\;\;\;\;\;\;(\rho_0)_{33}=\frac{1}{2}(\rho_0)_{22}\;.
\end{align}
The stationary density matrix enables us to calculate the emission intensity. Note that the harmonics with $n=\pm1$ do not contribute to the intensity 
\begin{align}
    I_1=\langle a^\dagger a\rangle&=\langle\Tr(\rho(t) a^\dagger a)\rangle_t\\ \nonumber
    =&\Tr(\rho_0 a^\dagger a)=(\rho_0)_{22}+2(\rho_0)_{33}\;.
\end{align}
This yields Eq.~\eqref{eq:I1} in the main text.
\subsection{Emission spectrum of a single emitter} \label{app:Spectrum}

We define emission spectrum as
\begin{equation}\label{eq:QReg0}
    S(\omega)=2\frac{\Omega}{2\pi}\Re\int_0^{2\pi/\Omega} dt'\int_0^\infty d\tau\; e^{-\rmi\omega \tau}\langle a^\dagger(t'+\tau)a(t')\rangle\;.
\end{equation}
The integration over $t'$ is performed over the period $2\pi/\Omega$ since the perturbation is periodic. According to the quantum regression theorem~\cite{Carmichael} 
\begin{equation}\label{eq:QRT}
    S(\omega)=2\frac{\Omega}{2\pi}\Re\int_0^{2\pi/\Omega}dt'\int_0^\infty d\tau\; e^{-\rmi\omega \tau}\Tr( a^\dagger \rho^a(t'+\tau))\;,
\end{equation}
where $\rho^a(t)$ satisfies Eq.~\eqref{eq:vector_rho} with the initial condition $\rho^a(t=t')=a \rho(t')$ which can be represented in the integral form as
\begin{align}
    \boldsymbol{\rho}^a(t'+\tau)&=e^{\mathcal{L}\tau}\boldsymbol{\rho}^a(t')\\ \nonumber
    &+\int_0^\tau d\tau' e^{\mathcal{L}(\tau-\tau')}\mathcal{F}(t'+\tau')\boldsymbol{\rho}^a(t'+\tau')\;,
\end{align}
where the perturbation $\mathcal{F}(t)=\frac{1}{2}\left(\mathcal{V} e^{\rmi\Omega t}-\mathcal{V}^*e^{-i\Omega t}\right)$ and initial $\rho^a(t')=a\rho_0 +a \rho_{-1} e^{\rmi \Omega t'}+a\rho_{1}e^{-\rmi\Omega t' }$. Then up to the second order in $g$, it reads  

\begin{align}
    &\boldsymbol{\rho}^a(t'+\tau)=e^{\mathcal{L}\tau}\boldsymbol{\rho}^a(t')\\ \nonumber
    &+\int_0^\tau d\tau' e^{\mathcal{L}(\tau-\tau')}\mathcal{F}(t'+\tau')e^{\mathcal{L}\tau'}\left(\boldsymbol{\rho}^a_{-1}e^{\rmi\Omega t'}+\boldsymbol{\rho}^a_1 e^{-\rmi\Omega t'}\right)\:.
\end{align}
Performing averaging over $t'$ and considering that $\mathcal{V}^*=-\mathcal{V}$ for a single emitter we get
\begin{align}   
    \langle\boldsymbol{\rho}^a(&t'+\tau)\rangle_{t'}=e^{\mathcal{L}\tau}\boldsymbol{\rho}_0^a\\ \nonumber&+\frac{1}{2}\int_0^\tau d\tau' e^{\mathcal{L}(\tau-\tau')}\mathcal{V}e^{\mathcal{L}\tau'}\left(\boldsymbol{\rho}^a_{-1}e^{-\rmi\Omega \tau'}+\boldsymbol{\rho}^a_{1}e^{\rmi\Omega \tau'}\right)\:.
\end{align}
This enable us to get the emission spectrum 
\begin{equation}\label{eq:spectr-master}
    S(\omega)=2\Re\int_0^\infty d\tau\; e^{-\rmi\omega \tau}\Tr( a^\dagger \langle\rho^a(t'+\tau)\rangle_{t'})\;.
\end{equation}
The result of the calculation according to Eq.~\eqref{eq:spectr-master} is shown in Fig.~\ref{fig:1-spectr}.
\section{Diagrammatic approach}\label{app:diagrams}

In case of weak modulation, only the states with small number of excitations are populated. They can be described in the framework of perturbative diagrammatic approach. The consideration generalizes the results  of Refs.~\cite{Fang2014,Poshakinskiy2016,Ke2019} and also Ref.~\cite{Sukhorukov2016} for the structures, modulated in time. Somewhat similar consideration for time-modulated structures is also available in Ref.~\cite{Ilin2022}, but that work does not consider parametric photon generation.

\begin{figure}
    \centering
    \includegraphics[width=0.7\columnwidth]{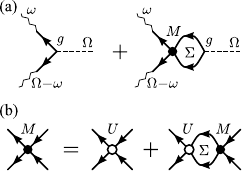}
    \caption{(a) Diagrams representing the amplitude of two-photon generation by modulated emitters. Dashed line represents the modulation, solid lines are the Green functions of emitter excitations, wavy lines are the outgoing photons. (b) The diagrammatic equation for the dressed vertex (solid circle) that describes the interaction of two emitter excitations. Open circle represents the bare vertex corresponding to the interaction amplitude $U$.}
    \label{fig:Adia}
\end{figure}

The diagram representing the (not normalized)
  wavefunction of the generated photon pair is shown in Fig.~\ref{fig:Adia}(a). 
The dashed line represents the modulation that creates a pair of emitter excitations (solid lines). Then, the excitations propagate in the structure, which is described by the Green functions (solid lines). The excitations can either get converted to the photons directly (first diagram) or interact and get converted to photons after that (second diagram). For the case when both photons are emitted in the left direction this yields
\begin{align}\label{eq:Apsi0}
\psi_{--}(\omega_1,\omega_2) = &\gamma_{\rm 1D} \sum_{ij} s_i^+(\omega_1)s_i^+(\omega_2) [1 + (M\Sigma)_{ij}] g_j \nonumber\\
& \times 2\pi\delta(\omega_1+\omega_2-\Omega) \,.
\end{align}
Here, the Green function of single excitation reads
\begin{align}
G(\omega) = (\omega-H)^{-1}, \qquad H_{ij}= \omega_0\delta_{ij} + D_{ij} \,,
\end{align}
the outer lines of the diagrams correspond to
\begin{align}
s_i^+(\omega) = \sum_j G_{ij}(\omega) \e^{\rmi q z_j} = \frac{\e^{\rmi q z_1}}{\rmi\gamma_{\rm 1D}} \left( \frac{\omega_0 - H}{\omega - H}\right)_{1i} \,,
\end{align}
and the propagation of a pair of excitations is described by
\begin{align}\label{eq:Asig}
\Sigma_{ij}(\Omega) = \rmi \int G_{ij}(\omega) G_{ij}(\Omega-\omega) \, \frac{d\omega}{2\pi} \,.
\end{align}
The dressed interaction vertex $M$ can be determined from the Dyson-like equation shown diagrammatically in  Fig.~\ref{fig:Adia}(b), which yields
\begin{align}
M(\Omega) = [U^{-1}-\Sigma(\Omega)]^{-1}\,.
\end{align}
Substituting this into Eq.~\eqref{eq:Apsi0} we finally obtain 
\begin{align}\label{eq:Apsif}
\psi_{--}(\omega_1,\omega_2) = &\frac{\gamma_{\rm 1D}}{U} \sum_{ij} s_i^+(\omega_1)s_i^+(\omega_2) M_{ij} g_j \nonumber\\
& \times 2\pi\delta(\omega_1+\omega_2-\Omega) \,.
\end{align}

The integration in Eq.~\eqref{eq:Asig} can be performed explicitly, which yields a compact expression for the $M$ matrix:
\begin{align}\label{eq:Amat}
M_{ij}(\Omega) = \left[\frac{ \Omega - H\otimes 1 - 1 \otimes H}{\Omega - H\otimes 1 - 1 \otimes H - \mathcal{U}}\,\mathcal{U}\right]_{ii,jj}
\end{align}
where $(A \otimes B)_{ij,kl} \equiv A_{ik}B_{jl}$, $\frac AB \equiv A B^{-1}$ and we introduced the diagonal $N^2 \times N^2$ matrix $ \mathcal{U}_{ij,kl} = \delta_{ij}\delta_{kl}\delta_{ik} U $.

\subsection{Correlation function}

The   second-order correlation function  of the emitted photon pair is calculated as
\begin{align}
G^{(2)}_{--}(\tau) = \frac{1}{\gamma_{\rm 1D}^2} \left| \int \psi_{--}(\omega_1,\omega_2) \e^{-\rmi\omega_1 t -\rmi\omega_2 (t+\tau )} \frac{d\omega_1 d\omega_2}{(2\pi)^2}\right|^2\:. 
\end{align}
At zero delay we use the representation Eq.~\eqref{eq:Amat} to obtain
\begin{align}\label{eq:AG20}
G^{(2)}_{--}(0) =\frac1{\gamma_{\rm 1D}^4} \left|  \sum_i \left[\frac{(\omega_0-H) \otimes (\omega_0-H)}{\Omega - H\otimes 1 - 1 \otimes H - \mathcal{U}}\right]_{11,ii} g_i\right|^2 
\end{align}

In the limit $U \gg \gamma_{\rm 1D}$ and for $|\Omega-2\omega_0-U| \lesssim \gamma_{\rm 1D}$, we keep in the denominator of Eq.~\eqref{eq:AG20} only the diagonal terms and find 
\begin{align}
G^{(2)}_{--}(0) 
= \dfrac{\left|  \sum\limits_i  g_i \,\e^{2\rmi q z_i} \right|^2}{(\Omega-2\omega_0-U)^2+4\gamma_{\rm 1D}^2 } \,,
\end{align}
where we used $ (\omega_0-H)_{1i} = \rmi\gamma_{\rm 1D} \e^{\rmi q(z_i-z_1)}$. 

For $|\Omega-2\omega_0| \sim \gamma_{\rm 1D}$, the result is 
\begin{align}
G^{(2)}_{--}(0) =\frac{1}{ U^2 } \left|  \sum_{ij} \Sigma^+_i(\Omega) [\Sigma^{-1}(\Omega)]_{ij} g_j \right|^2,  
\end{align}
where
\begin{align}
&\Sigma_{ij}(\Omega) = \left[ \frac1{\Omega - H\otimes 1 - 1 \otimes H} \right]_{ii,jj} , \\
&
\Sigma^+_i(\Omega) = \frac{1}{\gamma_{\rm 1D}^2}\left[\frac{(\omega_0-H) \otimes (\omega_0-H)}{\Omega - H\otimes 1 - 1 \otimes H}\right]_{11,ii} .
\end{align}

\subsection{Emission intensity}

We define the  intensity of emission to the left as
\begin{align}
I_-  = \int \left[ |\psi_{--}(\omega_1,\omega_2)|^2 + |\psi_{-+}(\omega_1,\omega_2)|^2 \right] \frac{ d\omega_1d\omega_2}{\gamma_{\rm 1D}T(2\pi)^2}\:,
\end{align}
where $T$ is the normalization time and $\psi_{-+}$ is the wave function of a pair of photons emitted in the opposite directions. The latter is obtained from Eq.~\eqref{eq:Apsi0} by replacing  the factor $s_j^+(\omega_2)$ with $s_j^-(\omega_2) = \sum_i G_{ij}(\omega) \e^{-\rmi q z_i}$.

Using the representation Eq.~\eqref{eq:Amat}, we get 
\begin{align}\nonumber
&\int |\psi_{--}(\omega_1,\omega_2)|^2  \tfrac{d\omega_1d\omega_2}{T(2\pi)^2} =  
\frac{1}{\rmi\gamma_{\rm 1D}^2} \sum_{ij}g_i g_j^* \\\Big[&\frac{(\omega_0-H_1)(\omega_0-H_2)(\omega_0-H_3^*)(\omega_0-H_4^*)}{(\Omega-H_1-H_4^*)} \\\times \nonumber
&\left(\frac{\Omega-H_3^*-H_4^*}{H_1-H_3^*} - \frac{\Omega-H_1-H_2}{H_4^*-H_2}\right) \\&\times
\frac1{(\Omega-H_1-H_2-\mathcal{U}_{12})(\Omega-H_3^*-H_4^*-\mathcal{U}_{34})}\Big]_{1111,iijj} \nonumber
\end{align}
and the corresponding integral of $|\psi_{-+}|^2 $ is obtained in the same way by taking the element $[\ldots]_{1N1N,iijj}$. Here $H_1 = H \otimes 1\otimes 1\otimes 1$, $H_2 =  1\otimes H \otimes1\otimes 1$ etc. 

In the limit $U \gg \gamma_{\rm 1D}$ and for $|\Omega-2\omega_0-U| \lesssim \gamma_{\rm 1D}$, the result can be  simplified:
\begin{align}
\int |&\psi_{--}(\omega_1,\omega_2)|^2  \tfrac{d\omega_1d\omega_2}{T(2\pi)^2}   =
\frac{2\rmi}{\gamma_{\rm 1D}^2[(\Omega-2\omega_0-U)^2+4\gamma_{\rm 1D}^2]} \nonumber\\&\times\sum_{ij}g_i g_j^* (\omega_0-H)_{1i}  (\omega_0-H^*)_{1j}{\mathcal Q}_{11,ij}\:, 
\\%
\int |&\psi_{-+}(\omega_1,\omega_2)|^2  \tfrac{d\omega_1d\omega_2}{T(2\pi)^2} =  \frac{\rmi}{\gamma_{\rm 1D}^2[(\Omega-2\omega_0-U)^2+4\gamma_{\rm 1D}^2]}  \nonumber\\\times 
&\sum_{ij}g_i g_j^* \Big\{
(\omega_0-H)_{1i}  (\omega_0-H^*)_{1j} \mathcal Q_{NN,ij}  \\+\nonumber
&(\omega_0-H)_{Ni}  (\omega_0-H^*)_{Nj} \mathcal Q_{11,ij} \Big\}\:,
\end{align}
where 
\begin{equation}
\mathcal Q=\frac{(\omega_0-H) \otimes (\omega_0 - H^*) }{(\omega_0-H) \otimes 1 -1 \otimes (\omega_0-H^*)}\:.
\end{equation}

\subsection{Optical theorem}
We assume that the modulation of the emitters is performed by an  external signal. The back-action of the emitter system on that signal can be accounted by introducing the scattering parameter $s$. 
Up to the second order in $g$, it reads
\begin{align}
&s = 1 - \rmi \sum_{ij} g_i^* \tilde\Sigma_{ij} g_j  , \\ &\tilde\Sigma_{ij}= \left[ \frac1{\Omega - H\otimes 1 - 1 \otimes H-\mathcal{U}} \right]_{ii,jj}\:.
\end{align}
In the absence of non-radiative losses, the optical theorem imposes $2|s|^2 + I_+ + I_- =2$. Therefore, we get
\begin{align}
&I_+ + I_- \\ \nonumber
&= -\frac{4}{\gamma_{\rm 1D}} {\rm Im\,} \sum_{ij} g_i^* g_j \left[ \frac1{\Omega - H\otimes 1 - 1 \otimes H-\mathcal{U}} \right]_{ii,jj} \:.
\end{align}
In the limit $U \gg \gamma_{\rm 1D}$ , $|\Omega-2\omega_0-U| \lesssim \gamma_{\rm 1D}$, the result is trivial
\begin{align}
I_+ + I_- =\frac{8}{(\Omega-2\omega_0-U)^2+4\gamma_{\rm 1D}^2}\, \sum_i |g_i|^2\:.
\end{align}

\subsection{Emission spectrum of a single emitter}

The emission spectrum can be calculated as 
\begin{equation}\label{eq:Spectr-Diag}
    S(\omega)=\int \frac{ |\psi_{--}|^2+ |\psi_{-+}|^2}{2\pi T\gamma_{\rm 1D}}d\omega'\;.
\end{equation}
In the case of a single emitter the wave function $\psi_{--}=\psi_{-+}$  and it can be found from Eq.~\eqref{eq:Apsif}
\begin{align}
    \psi_{--}(\omega,\omega')=&\frac{g\gamma_{\rm 1D}\left(\Omega-2\omega_0+2\rmi\gamma_{1\rm D}\right)}{\left(\Omega-2\omega_0-U+2\rmi\gamma_{\rm 1D}\right)}\\ \nonumber
    &\times\frac{2\pi \delta(\Omega-\omega-\omega')}{\left(\omega-\omega_0+\rmi\gamma_{\rm 1D}\right)\left(\omega'-\omega_0+\rmi\gamma_{\rm 1 D}\right)}\;.
\end{align}
The integration in Eq.~\eqref{eq:Spectr-Diag} yields Eq.~\eqref{eq:spectr} in the main text. 

\section{Two emitters}\label{app:two}

Here we apply the results of Appendix~\ref{app:diagrams}
for the system of two emitters. The general explicit expressions are quite bulky so consider two special cases. 

\subsection{$U \gg \gamma_{\rm 1D}$ and $|\Omega-2\omega_0-U| \lesssim \gamma_{\rm 1D}$}
In that limit,  we get
\begin{align}
G^{(2)}_{--}(0) =\frac{\left| g_1 + g_2 \e^{2\rmi q d}\right|^2}{(\Omega-2\omega_0-U)^2+4\gamma_{\rm 1D}^2} 
\end{align}
and
\begin{multline}
I_- =\frac{1}{(\Omega-2\omega_0-U)^2+4\gamma_{\rm 1D}^2} \\\times 
\frac{(7-3\cos 2qd) |g_1|^2 + (5-\cos 2qd)|g_2|^2 + 2\sin 2qd \,\text{Im\,}  g_1 g_2^*}{3-\cos 2qd}\:.\nonumber
\end{multline}
This yields Eqs.~\eqref{eq:Iminus} and \eqref{eq:g2mm} in the main text.

\subsection{Symmetric modulation, $g_1=g_2 =g$}
In that case, the two-photon emission is
\begin{align}\label{eq:g2symm}
&G^{(2)}_{--}(0) =4|g|^2\\&\times \left|\frac{(\Omega-2\omega_0) \cos qd + 2 \gamma_{\rm 1D} \sin qd}{(\Omega-2\omega_0+2\rmi\gamma_{\rm 1D})(\Omega-2\omega_0-U+2\rmi\gamma_{\rm 1D})+4\gamma_{\rm 1D}^2\e^{2\rmi q d}}\right|^2.\nonumber
\end{align}
Note that $G^{(2)}_{--}(0)$ vanishes at 
$
\Omega = 2\omega_0 -2 \gamma_{\rm 1D} \tan qd 
$.

The single-photon emission intensity is found easily from the optical theorem
\begin{align}
&I_+ = I_- = 8|g|^2\\\nonumber&\times \frac{(\Omega-2\omega_0+\gamma_{\rm 1D}\sin 2 qd)^2 +
2\gamma_{\rm 1D}^2 (3-\cos 2qd) \sin^2 qd}{\left|(\Omega-2\omega_0+2\rmi\gamma_{\rm 1D})(\Omega-2\omega_0-U+2\rmi\gamma_{\rm 1D})+4\gamma_{\rm 1D}^2\e^{2\rmi q d}\right|^2}\:.
\end{align}
We note that $I_{\pm}$ has a minimum at $\Omega \approx 2\omega_0 - \gamma_{\rm 1D} \sin 2 qd$.  In particular, for $qd =0$, we have $I_\pm(\Omega = 2\omega_0) = 0$.

\bibliography{casimir}
\end{document}